\documentclass[aps,prl,showpacs,showkeys,twocolumn,groupedaddress,preprintnumbers]{revtex4}
\usepackage{graphicx}
\usepackage{dcolumn}
\usepackage{bm}
\usepackage{psfrag}
\usepackage{amssymb}

\begin{document}
\preprint{LMU-ASC 27/06}
\preprint{MPP-2006-39}
\title{Escaping from the black hole?}
\author{E. Babichev$^{1}$, V. Mukhanov$^{2}$ and A. Vikman$^{2}$}
\affiliation{$^{1}$Max-Planck-Institut f\"ur Physik, F\"ohringer Ring 6,
D-80805, Munich,
Germany \\
and Institute for Nuclear Research of the Russian Academy of Sciences, 60th
October Anniversary Prospect 7a, 117312 Moscow, Russia\\
$^{2}$Arnold-Sommerfeld-Center for Theoretical Physics, Department f\"ur
Physik, Ludwig-Maximilians-Universit\"at M\"unchen, Theresienstr. 37,
D-80333, Munich, Germany}

\begin{abstract}
We show that if there exists a special kind of Born-Infeld type scalar
field, then one can send information from inside a black hole. This
information is encoded in perturbations of the field propagating in
non-trivial scalar field backgrounds, which serves as a "new ether". Although
the theory is Lorentz-invariant it allows, nevertheless, the superluminal
propagation of perturbations with respect to the "new ether". We found the
stationary solution for background, which describes the accretion of the
scalar field onto a black hole. Examining the propagation of small
perturbations around this solution we show the signals emitted inside the
horizon can reach an observer located outside the black hole. We discuss
possible physical consequences of this result.
\end{abstract}


\maketitle

\emph{Introduction.}-
%
During last years the scalar fields, described by the Lagrangians with a
non-standard kinetic term, attracted a considerable interest. Such
structures are rather common for effective fields theories.
In cosmology they were first introduced in the context of k-inflation \cite%
{k-inflation} and then the k-essence models were suggested for solving the
cosmic coincidence problem\cite{k-essence}. Tachyon matter \cite{tachyon},
ghost condensate \cite{ghost} and phantom \cite{phantom} can be thought as
the further developments of this idea.

In some cases Lorentz invariant theories with nonlinear kinetic terms allow
the superluminal propagation of perturbations on dynamical backgrounds and
this may have interesting applications in cosmology \cite{MukhGar,MukhVik}%
. We would like to point out that the issue of causality is rather
nontrivial in the theories with superluminal propagation and requires
further investigation. For example, the Cauchy problem is well-posed not for
all initial data \cite{ArmenLim, Rendall, ArkHamDubov,HashItz}.

One of the interesting issues is the behavior of noncanonical scalar fields in
the neighborhood of black holes \cite{Mukohu,BDE,Bron} and in this paper we
investigate the consequences of the superluminal propagation of such fields 
in the black hole background. 
In particular, we will consider a Lorentz invariant scalar
field theory with Lagrangian which allows the superluminal propagation of
perturbations during accretion onto black hole. Assuming that the
backreaction of the scalar field on the metric is negligible we will find
first the analytic solution describing the spherically symmetric accretion
of the scalar field.
After that we investigate the propagation of the perturbations in this
background.

\emph{Model.-}Let us consider a scalar field with the action
\begin{equation}
S_{\phi }=\int d^{4}x\sqrt{-g}p(X),
\end{equation}%
where the Lagrangian density is given by
\begin{equation}
p(X)=\alpha ^{2}\left[ \sqrt{1+\frac{2X}{\alpha ^{2}}}-1\right] -\Lambda .
\label{Lagrange}
\end{equation}%
It depends only on $X\equiv \frac{1}{2}\nabla _{\mu }\phi \nabla ^{\mu }\phi
$, and $\alpha $ and $\Lambda $ are free parameters of the theory.
Throughout the paper $\nabla _{\mu }$ denotes the covariant derivative and
we use the natural units in which $G=\hbar =c=1$. \ The kinetic part of the
action is the same as in \cite{MukhVik} and for small derivatives, that is,
in the limit $2X\ll \alpha ^{2},$ it describes the usual massless free
scalar field. One can prove that the theory, described by (\ref{Lagrange})
is ghost-free.

The equation of motion for the scalar field is
\begin{equation}
G^{\mu \nu }\nabla _{\mu }\nabla _{\nu }\phi =0,  \label{general eom}
\end{equation}%
where the induced metric $G^{\mu \nu }$ is given by
\begin{equation}
G^{\mu \nu }\equiv p_{,X}g^{\mu \nu }+p_{,XX}\nabla ^{\mu }\phi \nabla ^{\nu
}\phi,  \label{G}
\end{equation}
and $p_{,X}\equiv \partial p/\partial X$. This equation is hyperbolic and
its solutions are stable with respect to high frequency perturbations
provided $(1+2Xp_{,XX}/p_{,X})>0$ \cite{MukhGar,ArmenLim,Rendall}. This
condition is always satisfied in the model under consideration. It is well
known that, if $\nabla _{\nu }\phi $ is timelike (that is, $X>0$ in our
convention), then the field described by (\ref{Lagrange}) is formally
equivalent to a perfect fluid with the energy density $\varepsilon
(X)=2Xp_{,X}(X)-p(X)$, the pressure $p=p(X)$ and the four-velocity
\begin{equation}
u_{\mu }=\frac{\nabla _{\mu }\phi }{\sqrt{2X}}.  \label{4velocity}
\end{equation}
%
The effective sound speed of perturbations is given by
\begin{equation}
c_{s}^{2}\equiv \frac{\partial p}{\partial \varepsilon }=1+\frac{2X}{\alpha
^{2}}.  \label{soundspeed}
\end{equation}%
and for $X>0$ it always exceeds the speed of light. For the further
considerations it occurs to be convenient to express the energy density and
pressure in terms of this speed of sound, namely,
\begin{equation}
\varepsilon =\alpha ^{2}(1-c_{s}^{-1})+\Lambda ,~~p=\alpha
^{2}(c_{s}-1)-\Lambda .  \label{enpres_c}
\end{equation}%
It is easy to see that the Null Energy Condition is valid and hence the black hole
area theorem \cite{Hawking} holds.

\emph{Background solution.-}
%
\begin{figure}[t]
\psfrag{r}[t]{$r/r_{g}$} \psfrag{x}[t]{$x_{*}$} \psfrag{c}[b]{%
\large$c_{s}^{2}$} \psfrag{E}[b]{\Large$\frac{\varepsilon-\Lambda}{%
\alpha^{2}}$} \includegraphics[width=0.48\textwidth]{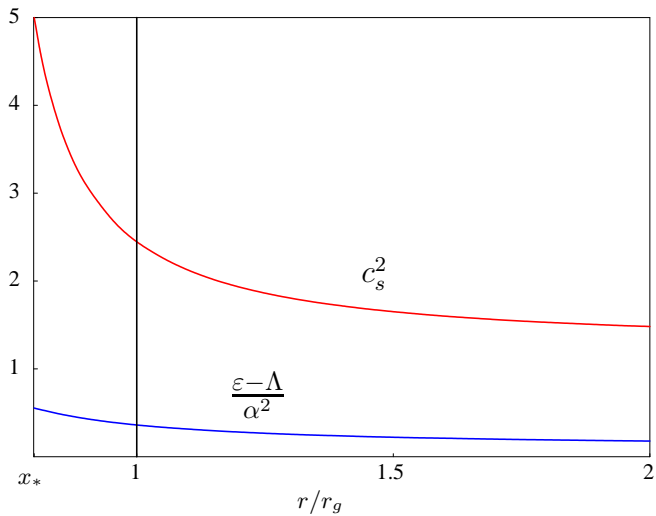}
\caption{For the background solution in the case $c_{\infty }^{2}=5/4$ the
squared sound speed (red) and the normalized energy density, $(\protect%
\varepsilon -\Lambda )/\protect\alpha ^{2}$, (blue) are shown as functions
of radial coordinate $x\equiv r/r_{g}$. The sound horizon $r_{\ast }/r_{g}=4/5$
is located inside the Schwarzschild horizon.}
\label{fig1}
\end{figure}
First we will find a stationary spherically symmetric background solution
for the scalar field falling onto a Schwarzschild black hole. To describe
the black hole we use the ingoing Eddington-Finkelstein coordinates, in
which the metric takes the form:
\begin{equation}
ds^{2}=f(r)dV^{2}-2dVdr-r^{2}d\Omega ,  \label{metric}
\end{equation}%
where $f(r)\equiv 1-r_{g}/r$ and $r_{g}\equiv 2M$ is the gravitational radius of the
black hole. The coordinate $V$ is related to the Schwarzschild coordinates $t
$ and $r$ as: $V\equiv t+r+r_{g}\ln |r/r_{g}-1|$. Let us assume that the
infalling field has a negligible influence on the metric, that is, we
consider an accretion of the test fluid in the given gravitational field. 
The requirement of stationarity implies the following ansatz for the solution:
\begin{equation}
\phi (V,x)=\alpha \sqrt{c_{\infty }^{2}-1}\left( V+r_{g}\int F(x)dx\right) ,
\label{anz}
\end{equation}%
where $x\equiv r/r_{g}$ and $c_{\infty }$ is the speed of sound at infinity. The
overall factor in (\ref{anz}) is chosen to recover the cosmological solution
at infinity: $\phi (V,x)\rightarrow \alpha t\sqrt{c_{\infty }^{2}-1}$ and $%
r_{g}$ in front of the integral is for the further convenience. The function $%
F(x)$ must be determined solving equations of motion for appropriate
boundary conditions. Substituting (\ref{anz}) into (\ref{general eom}) and
integrating over $r$ we obtain the following equation for the function $F(x)$%
:
\begin{equation}
\frac{(fF+1)x^{2}}{\sqrt{1-\left( fF^{2}+2F\right) \left( c_{\infty
}^{2}-1\right) }}=\frac{B}{c_{\infty }^{4}},  \label{eq F}
\end{equation}%
where $B$ is the constant of integration. The solution of (\ref{eq F}),
which is nonsingular at the black hole horizon, is given by:
\begin{equation}
F(x)=\frac{1}{f}\left( B\sqrt{\frac{c_{\infty }^{2}+f-1}{fx^{4}c_{\infty
}^{8}+B^{2}\left( c_{\infty }^{2}-1\right) }}-1\right) .  \label{F}
\end{equation}%
The speed of sound speed can then be found using (\ref{F}), (\ref{anz}) and (%
\ref{soundspeed}):
\begin{equation}
c_{s}^{2}=\frac{x^{3}c_{\infty }^{8}\left( xc_{\infty }^{2}-1\right) }{%
(x-1)x^{3}c_{\infty }^{8}+B^{2}\left( c_{\infty }^{2}-1\right) }.  \label{cs}
\end{equation}%
Note that the speed of sound becomes infinite at some $x\equiv x_{sing}$ and
this singularity is physical if the real regular solution (\ref{F}) exists
for all $x>x_{sing}$.

A constant of integration $B,$ entering (\ref{F}) and (\ref{cs}), determines
the energy flux falling onto the black hole. To fix it we have to find the
solution which is non-singular on the \textit{sound horizon }and outside it.
Below we consider the propagation of perturbations and find how the position
of the sound horizon depend on $B$. Then, given $c_{\infty },$ and comparing
the positions of the singularities and the sound horizon we determine the
unique value for $B$.

\emph{Small perturbations.-}
%
Let us now consider the small perturbations around background (\ref{anz}), (%
\ref{F}). The characteristics (propagation vectors $\eta ^{\mu }$) for
equation (\ref{general eom}) satisfy the following equation (see, e.g. \cite%
{ArmenLim,Rendall}):
\begin{equation}
\tilde{G}_{\mu \nu }\eta ^{\mu }\eta ^{\nu }=0,  \label{null}
\end{equation}
where $\tilde{G}_{\mu \nu }$ is the matrix inverse to $G^{\mu \nu },$ that
is, $G^{\mu \nu }\tilde{G}_{\nu \sigma }=\delta _{\sigma }^{\mu }{},$ and it is
calculated for the background solution (\ref{anz},\ref{F}). The vector $\eta
^{\mu }$ describes the propagation of the wave front. After lengthy, but
straightforward calculations, we obtain from (\ref{null}) and (\ref{G}) the
following differential equation for the characteristics $\eta _{\pm
}(x)\equiv dV/dx$:
\begin{equation}
\eta _{\pm }=\frac{1}{f}+\frac{1}{\xi _{\pm }},  \label{eta}
\end{equation}%
where
\begin{equation}
\xi _{\pm }=\pm f\sqrt{c_{\infty }^{2}-\frac{1}{x}}\,\frac{\sqrt{%
B^{2}(c_{\infty }^{2}-1)+c_{\infty }^{8}x^{4}f}}{c_{\infty }^{4}x^{2}f\mp
B(c_{\infty }^{2}-1)}.  \label{ksi}
\end{equation}%
It is worth mentioning that the equation $\xi _{\pm }=dx/dt$ determines the
propagation of wave front in the Schwarzschild coordinates $x$ and $t$.

\begin{figure}[t]
\psfrag{V}[r]{\large$V\over r_{g}$} \psfrag{R}[t]{$r/r_{g}$} %
\psfrag{x}[t]{\small$x_{*}$} \includegraphics[width=0.45%
\textwidth,height=200pt]{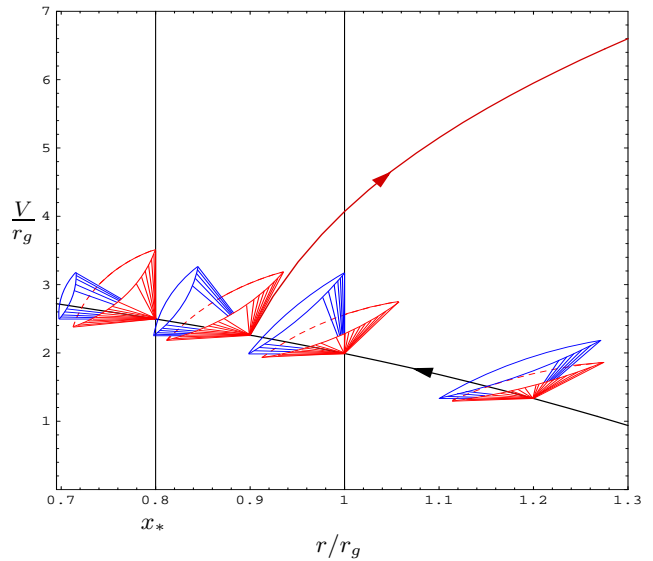}
\caption{In the Eddington-Finkelstein coordinates the emission of a 
sound signal from the falling spacecraft is shown. The blue cones correspond 
to the future light cones and the red cones are the future sonic cones 
(\protect\ref{eta}). The black curve represents the world line obtained 
numerically from Eqs. (\protect\ref{4velocity}), (\protect\ref{anz}), 
(\protect\ref{F}) for the
spacecraft which moves together with a falling background field. Being
between the the Schwarzschild ($r=r_{g}$) and sound ($r=r_{\ast }$) horizons
the spacecraft emits an acoustic signal (shown by red) which reaches the
distant observer in finite time. The trajectory of the signal is obtained by
the numerical integration of Eq.~(\protect\ref{eta}).}
\label{cones}
\end{figure}

Equation (\ref{eta}) does not specify the \textit{direction} of the
propagation completely. In addition to the value of $dV/dx$ one has to
choose a cone of \textit{future} and a cone of \textit{past} for every event.
However, the position of the past and the future lightcones helps us to fix
the past and the future cones for the scalar field perturbations, or in
other words, for the "sound". Using characteristics (\ref{eta}) we then
select uniquely the sonic cones as follows: i) the past and the future sonic
cones should not have overlapping regions; ii) the future sonic cone
contains the future light cone, while the past sonic cone contains the past
light cone. This last property can be justified because it holds at the
spatial infinity and the sonic characteristics (\ref{eta}) nowhere coincide
with the radial light geodesics (otherwise for the sonic signal $ds^{2}$
would vanish somewhere and this is obviously not true). As a result we
conclude: a signal propagating along $\eta _{+}$ points in the positive $V-$%
direction, while a signal corresponding to $\eta _{-}$ points in the
negative $V-$direction (see Fig.~\ref{cones}).

Having calculated the propagation vectors we can find the \textit{sonic
horizon}. The \textit{sonic horizon} is defined as a surface, where the
length of the spatial velocity vector is equal to the speed of sound.
Outside this surface the signals can reach the spatial infinity, while sound
cannot escape from inside because it is trapped by the supersonic motion of
a fluid (in the same way as light is trapped inside the event horizon by the
gravitational field). The acoustic signal directed \textit{out} of the black
hole corresponds to $\eta _{+}$ and therefore the \textit{sound horizon} is
located at $x\equiv x_{\ast }$ where $\eta _{+}\equiv (dV/dx)_{+}$ becomes
infinite (see Fig.~\ref{cones}).
Now we can fix a constant of integration $B,$ entering (\ref{anz}), (\ref{F}%
). We simply demand that in the physically occurring situation there exists
no singularity on the \textit{sound horizon }and outside of it. This
procedure is similar to that one arising in the problem of perfect fluid
accretion where the physical solution does not diverge at the event horizon
(see, e.g.~\cite{Beskin}). Thus, fixing $B$ reduces to the analysis of the
mutual location of $x_{sing}$ and $x_{\ast }$. After some calculations we
find the following:

\begin{itemize}
\item For $B\neq 1$ either the physical singularity coincides with the sound
horizon or the speed of sound becomes imaginary (this means absolute
instability) within some region outside the singular surface, for $%
x>x_{sing} $. In both cases the solution is nonphysical.

\item For $B=1$ and $c_{\infty }^{2}>4/3$ the speed of sound becomes
imaginary before reaching of sound horizon or singularity. This solution is
also nonphysical.

\item For $B=1$ and $c_{\infty }^{2}<4/3$ the sound horizon is located at $%
x_{\ast }=1/c_{\infty }^{2}$ and the singularity is hidden inside the sound
horizon, $x_{sing}<x_{\ast }$. This is the only physically relevant solution
we are searching for.
\end{itemize}

Thus, we have to set $B=1$ in (\ref{anz}), (\ref{F}) and this ends the
constructing of the background.

Before we turn to the discussion of the signals propagation we will briefly
analyze the validity of the stationarity approximation when the backreaction
can be neglected. Having fixed $B$ the rate of the accretion can be easily
evaluated as (see e.g. \cite{BDE}):
\begin{equation}
\dot{M}=4\pi M^{2}\alpha ^{2}(c_{\infty }^{2}-1)/c_{\infty }^{4}.
\label{dotM}
\end{equation}%
It is clear that for any fixed value of $c_{\infty }$ we can choose a small
enough $\alpha$, so that the energy flux onto black hole is negligible. The
propagation of perturbations (\ref{eta}) on the background (\ref{anz}) does
not depend on $\alpha $, but only on $c_{\infty }$. Therefore, we can always
take sufficiently small $\alpha $ in (\ref{Lagrange}) to ensure that during
the gedanken experiment with sending signals from the interior of a black
hole the background solution remains nearly unchanged.

After we have found the physically relevant background solution we will
discuss whether the acoustic signals can really escape from the interior of
the black hole. This becomes possible because in the case under
consideration the sound horizon ($x_{\ast }=1/c_{\infty }^{2}$) is located
inside the Schwarzschild radius. As long as the signals are emitted at large
enough $x$, namely, at $x>x_{\ast },$ they reach the spatial infinity
propagating along $\eta _{+}$. For example, at the event horizon we have:
\begin{equation}
\eta _{\pm H}=\frac{1}{2}\frac{\left( c_{\infty }^{4}\pm 1\right) ^{2}}{%
c_{\infty }^{2}-1}.
\end{equation}%
The propagation vector $\eta _{+H}$ is positive and so signals could freely
penetrate the Schwarzschild horizon and move \textit{outside} the black hole.
The Fig.~\ref{cones} shows how the acoustic signals go out from the interior
of a black hole.

Let us calculate the redshift of the emitted signal. Suppose that a
spacecraft moves together with the falling background field (such that in
the spacecraft's system of coordinates $\nabla \phi _{0}=0$) and sends the
acoustic signals with the frequency $\omega _{0}$. After a simple
geometrical exercise in the plane $(V,x)$ one can obtain that an observer at
rest at the spatial infinity will detect these signals at the frequency:
\begin{equation}
\omega _{\infty }=\omega _{0}\frac{c_{\infty }^{8}x^{3}(x-1)+c_{\infty
}^{2}-1}{x^{2}c_{\infty }^{4}(x^{2}c_{\infty }^{4}+1)}.  \label{redshift}
\end{equation}%
Note that the ratio $\omega _{\infty }/\omega _{0}$ is finite for any $%
x>x_{\ast }$ and it vanishes for $x=x_{\ast }$. 

\emph{Conclusion.-}
%
The main result of this paper can be summarized as follows: if there exist a
specific Born-Infeld type fields, then during accretion of these fields onto
black hole one can send information from the interior of the black hole. We
would like to stress that this result has a classical origin and no quantum
phenomena are involved. The discussed effect changes the universal meaning
of the Schwarzschild horizon as an event horizon and may have important
consequences for the thermodynamics of black holes.

We consider the present work as simply an illustration of a concept. The
particular theory examined does not have any justification from the 
point of view of particle physics. However, for a wide class of 
nonlinear theories the
situation can be similar and therefore it is quite possible that the
information can really be send from inside the black hole.

Also we would like to point out that in our model the cosmic censorship
hypothesis is holds because the singularity is hidden by the sound horizon.
The null energy condition is not violated as well. Hence the Schwarzschild
horizon never decreases.

Note: The recent paper \cite{DubovSib} deals with thermodynamics of black
holes in the presence of superluminal fields. However, the model analyzed in
this paper is completely different from ours, namely, the authors of \cite%
{DubovSib} have considered two kinetically coupled fields, one of which is
the ghost condensate \cite{ghost}.
Moreover, the similar possibility of sending signals from the inside of a black hole opens in bigravity theories \cite{Bi}. 

We are grateful to S.~Winitzki and A. Rendall 
for very helpful discussions and V.~Beskin for
useful correspondence. The work of E.B. was supported by Alexander von
Humboldt fellowship grant; V.M. thanks SFB 375 for support.

\end{document}